\documentclass[prc,showpacs,showkeys,superscriptaddress,nofootinbib,floatfix,twocolumn,groupedaddress]{revtex4}
\usepackage{color}
\usepackage{amsfonts}
\usepackage{amsbsy}
\usepackage{mathrsfs}
\usepackage{graphicx}
\usepackage{bm}
\def\lsim{\mathrel{\rlap{
\lower4pt\hbox{\hskip-3pt$\sim$}}
    \raise1pt\hbox{$<$}}}     
\def\gsim{\mathrel{\rlap{
\lower4pt\hbox{\hskip-3pt$\sim$}}
    \raise1pt\hbox{$>$}}}     
\def\scr#1{\mbox{\scriptsize #1}}

\newcommand{\be}{\begin{equation}}
\newcommand{\ee}{\end{equation}}
\newcommand{\bel}[1]{\be\label{#1}}
\newcommand{\hsp}{\hspace*{1pt}}

\begin{document}
\title{Elliptic Flow and Dissipation in Heavy-Ion Collisions at %
$E_{\scr{lab}}\simeq$ (1--160)$A$ GeV%
\footnote{This
    paper is dedicated to our long-term coauthor and friend,
    V.N.~Russkikh, without whom this and many other works would be impossible.}
}
\author{Yu.B.~Ivanov}\thanks{e-mail: Y.Ivanov@gsi.de}
\affiliation{Frankfurt Institute for Advanced Studies, J.W. Goethe Universit\"at,
Ruth-Moufang-Str. 1, D60438 Frankfurt am Main, Germany}
\affiliation{The Kurchatov Institute, Acad. Kurchatov
Sq. 1, Moscow 123182, Russia}
\author{I.N. Mishustin}
\affiliation{Frankfurt Institute for Advanced Studies, J.W. Goethe Universit\"at,
Ruth-Moufang-Str. 1, D60438 Frankfurt am Main, Germany}
\affiliation{The Kurchatov Institute, Acad. Kurchatov
Sq. 1, Moscow 123182, Russia}
\author{\fbox{V.N.~Russkikh}}
\affiliation{The Kurchatov Institute, Acad. Kurchatov
Sq. 1, Moscow 123182, Russia}
\author{L.M. Satarov}
\affiliation{Frankfurt Institute for Advanced Studies, J.W. Goethe Universit\"at,
Ruth-Moufang-Str. 1, D60438 Frankfurt am Main, Germany}
\affiliation{The Kurchatov Institute, Acad. Kurchatov
Sq. 1, Moscow 123182, Russia}
\begin{abstract}
Elliptic flow in heavy-ion collisions at incident energies
$E_{\scr{lab}}\simeq$ (1--160)$A$ GeV is analyzed within the model of 3-fluid
dynamics (3FD). We show that a simple correction factor,
taking into account dissipative affects,
allows us to adjust the 3FD results to
experimental data. This single-parameter fit results in a good
reproduction of the elliptic flow as a function of the incident energy,
centrality of the collision and rapidity.
The experimental scaling of pion eccentricity-scaled elliptic
flow versus charged-hadron-multiplicity density
per unit transverse area turns out to be also reasonably described.
Proceeding from values of the Knudsen number, deduced from this fit,
we estimate the upper limit the shear viscosity-to-entropy ratio
as $\eta/s \sim 1-2$
at the SPS incident energies. This value is of the order of minimal $\eta/s$
observed in water and liquid nitrogen.

\end{abstract}
%

\maketitle

\section{Introduction}

The elliptic flow ($v_2$) of produced particles is one of the most
sensitive observables which brings information about the
degree of collectivity during the expansion stage of heavy-ion collisions
\cite{Ollitrault92,Voloshin96,Voloshin08}. When the collectivity is strong, like in
the case of ideal hydrodynamics, the elliptic flow takes the highest value
(the so called hydrodynamic limit). If the
collectivity is weak, like in a dilute system of weakly interacting particles, it is close
to zero. Therefore, it is not surprising that~$v_2$ is highly sensitive
to dissipative effects which are associated with attenuation of the collectivity
during the expansion stage. The elliptic flow turns out to be considerably reduced
by the cascade ''afterburner'' following the hydrodynamic freeze-out
\cite{Teaney01,Hirano07,Teaney:2003kp} as well as by viscosity effects
\cite{Romatschke:2007mq,Luzum:2008cw,Song:2007fn,Song:2007ux,Dusling:2007gi,Song:2008si}.
Note that the afterburner can be considered as a strong viscosity effect at the
final stage of the fireball expansion.

All above mentioned analyses of the dissipative effects were carried out for
high incident energies in the RHIC region. In this paper we would like to
turn to lower energies of the

SIS--AGS--SPS region. The experimental data
\cite{E895-02,E877-v2,FOPI05,FOPI07,NA49-98-v1,NA49-99,NA49-03-v1,CERES:INPC01}

in this energy region are much more fragmentary than in the RHIC domain.
In the future they will be essentially complemented by new
facilities, FAIR in Darmstadt and NICA in Dubna, as well as by experiments within
the low-energy-scan program at RHIC. The available data were analyzed within kinetic
\cite{Fuchs,Sahu02,IOSN05,SBBSZ04} and 3FD \cite{3FDflow,3FD-GSI07} models.
It was found that the 3FD approach noticeably overestimates
the data at the SPS energies independently of the stiffness of the equation of state
(EoS) and stopping power used by the model. Recently the SPS data were studied within
a hybrid hydrocascade model~\cite{Bleicher09} which includes afterburner
effects. It was demonstrated that the afterburner
indeed essentially reduces the hydrodynamic $v_2$ values which, however, still exceed
their experimental values.

In fact, there were no studies of dissipative effects related
to the elliptic flow in the AGS--SPS energy range.  In this paper we would like
to present such a study based on simulations within the 3FD model
\cite{3FDflow,3FD-GSI07,3FD}.  In order to estimate the dissipation,
we use the
approach suggested in Refs.~\mbox{\cite{Dumitru07,Ollitrault05,Ollitrault08}}.
We also study scaling of the elliptic flow with the midrapidity density of
produced charged particles, which was experimentally revealed in Refs. \cite{Voloshin08,Voloshin00}.

\section{3FD Model}

In Ref. \cite{3FD} we have introduced a 3-fluid
dynamical model for simulating heavy-ion
collisions in the c.m. energy range $\sqrt{s_{NN}}= 2.3-30$
GeV\hsp\footnote{
The upper limit is associated with computational demands of
the 3FD code. The lower limit cuts off the region where
applicability of the hydrodynamics becomes questionable.
}
(or $E_{\rm lab}= (1-500)A$ GeV in terms of lab. energy of the beam),
which overlaps with the
SIS--AGS--SPS
energy range
and covers the domains of future FAIR and NICA facilities.
The 3FD model is a
straightforward extension of the 2-fluid model with radiation of
direct pions \cite{MRS88,gsi94} and the (2+1)-fluid model
\cite{Kat93}. These models were extended in such a
way that the created baryon-free fluid (a
``fireball'' fluid) is treated on equal footing with the baryon-rich fluids.
For the fireball fluid we have introduced a certain formation time, during
which it evolves without interaction with other fluids.

\begin{figure}[b]
%
\includegraphics[width=6cm]{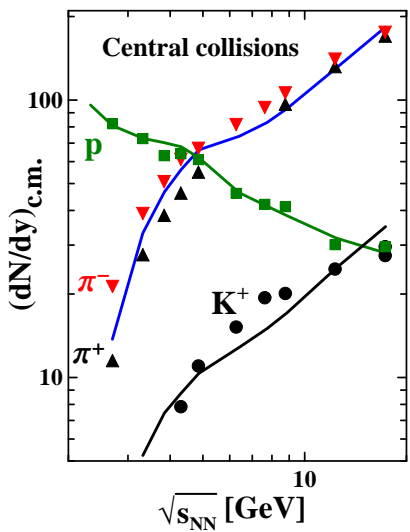}
$\;$\vspace*{-12mm}
\caption{(Color online) Incident energy dependence of midrapidity
  yield of various charged
  hadrons produced in central Au+Au and Pb+Pb collisions.
The 3FD calculations are done with the hadronic EoS ($K=$ 190 MeV).
The compilation of experimental data is taken from Ref. \cite{Andronic}.}
\label{fig1}
\end{figure}

\begin{figure*}[htb]
$\;$\vspace*{-22mm}

\includegraphics[width=12.5cm]{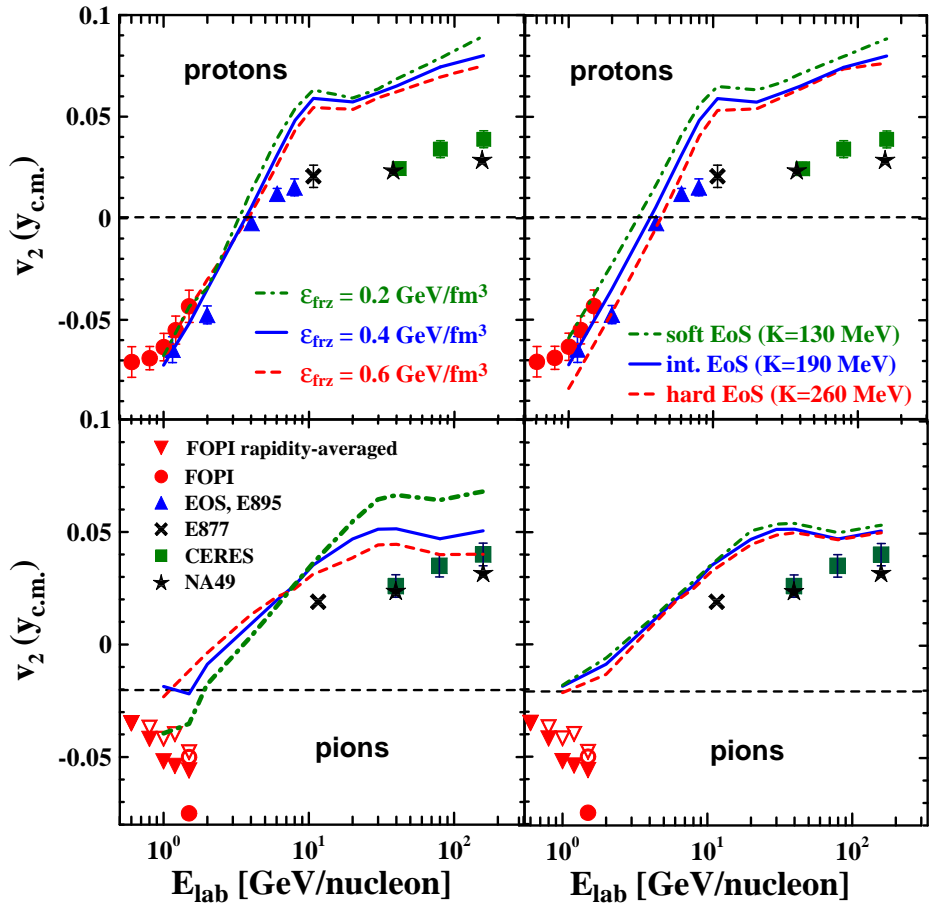}
$\;$\vspace*{-28mm}
\caption{(Color online)
Elliptic flow of protons (top panels) and pions (bottom panels)
at midrapidity as a function of incident energy
in mid-central
Au+Au (at SIS and AGS energies) and Pb+Pb (at SPS energies) collisions.
The 3FD calculations with soft ($K=130$ MeV), intermediate ($K=190$ MeV)
and hard  ($K=260$ MeV) EoS's at standard freeze-out energy density
$\varepsilon_{\rm frz}=0.4$ GeV/fm$^3$ are displayed in right panels.
Calculations with intermediate ($K=190$ MeV)
EoS and different freeze-out energy densities
$\varepsilon_{\rm frz}=$ 0.2, 0.4 and 0.6 GeV/fm$^3$ are presented in left panels.
Compilation of experimental data is from Ref.~\cite{FOPI05}.
The FOPI pion data are from \cite{FOPI07}: filled symbols correspond
to positive pions, and open  symbols, to negative pions;
the rapidity average is taken over the interval
$ -1.8 < y-y_{\rm c.m.} < 0$.
}
\label{fig2}
\end{figure*}

The input required by the model consists of the EoS and inter-fluid
friction forces. Our goal is to find an EoS which is able to reproduce
in the best way the largest body of  available observables. The friction forces
determine the stopping power of colliding nuclei  and thereby the rate of
thermalization of produced matter. In principle, the friction
forces  and the EoS are not independent, because medium modifications, providing a
nontrivial EoS, also modify elementary cross sections. However, currently
we have at our disposal only a rough estimate of the inter-fluid
friction forces~\cite{Sat90}. In the present version of the 3FD model these
forces are fitted to the stopping power observed in proton rapidity
distributions.

We have started our simulations~\cite{3FD}
with a simple, purely hadronic EoS \cite{gasEOS} which involves only
a density dependent mean-field providing saturation of cold
nuclear matter at normal nuclear density $n_0=$ 0.15 fm$^{-3}$
with the proper binding energy $-16$ MeV and a certain incompessibility~$K$.
It includes 48 lowest mass hadronic states~\cite{3FD}.
This EoS is a natural reference point for any other more elaborated~EoS.

The 3FD model turns out to reasonably
reproduce a great body of experimental data in a wide
energy range from AGS to SPS. Figure \ref{fig1} illustrates the results of Ref. \cite{3FD}.
Here we use the compilation~\cite{Andronic} of experimental data
from Refs.~\cite{E877:piKp,E895:piKp,E917:piKp,NA49:piKp,NA44:piKp,NA57:piKp}.
These data slightly differ in degree of centrality.
For Au+Au collisions at AGS energies and Pb+Pb reactions at $E_{\rm lab}=158A$ GeV
we perform our calculations taking the fixed impact  parameter $b=$ 2 fm.
For Pb+Pb collisions at lower SPS energies we use  $b=$ 2.5 fm.
The overall description of the data is quite good in the whole energy
range under consideration.
However, a certain underestimation of the kaon yield at lower SPS energies
prevents the model from reproducing the ''horn'' in the $K^+/\pi^+$
ratio observed experimentally in Refs.~\cite{gaz,gaz2}.


In the present simulations we use slightly different set of parameters
as compared with that of Ref. \cite{3FD}. This is caused by several
reasons. We excluded all contributions of weak decays into hadronic
yields, as it required by the NA49 experimental data. Strong decays of
baryon resonances were updated. Scalar meson~$f_0(600)$ was included in the list
of produced mesons. Some bugs in the code were  corrected.
All this inspired a refit of the model parameters. The best EoS, which
reproduces the main body of experimental data, corresponds now to
$K=190$ MeV (instead of $K=210$ MeV in Ref. \cite{3FD}).
In fact, the EoS's with $K=$ 190 and 210 MeV are very close to each
other. The enhancement factor of the friction forces estimated from the
proton-proton cross sections was reduced: we take the coefficient
$\beta_h=0.5$ instead of $\beta_h=0.75$ in Eq. (39) of
Ref.~\cite{3FD}. With this updated set of parameters the reproduction
of available data is approximately the same as in~\cite{3FD}.

\begin{figure}[thb]
\hspace*{-7mm}
\includegraphics[width=9.5cm]{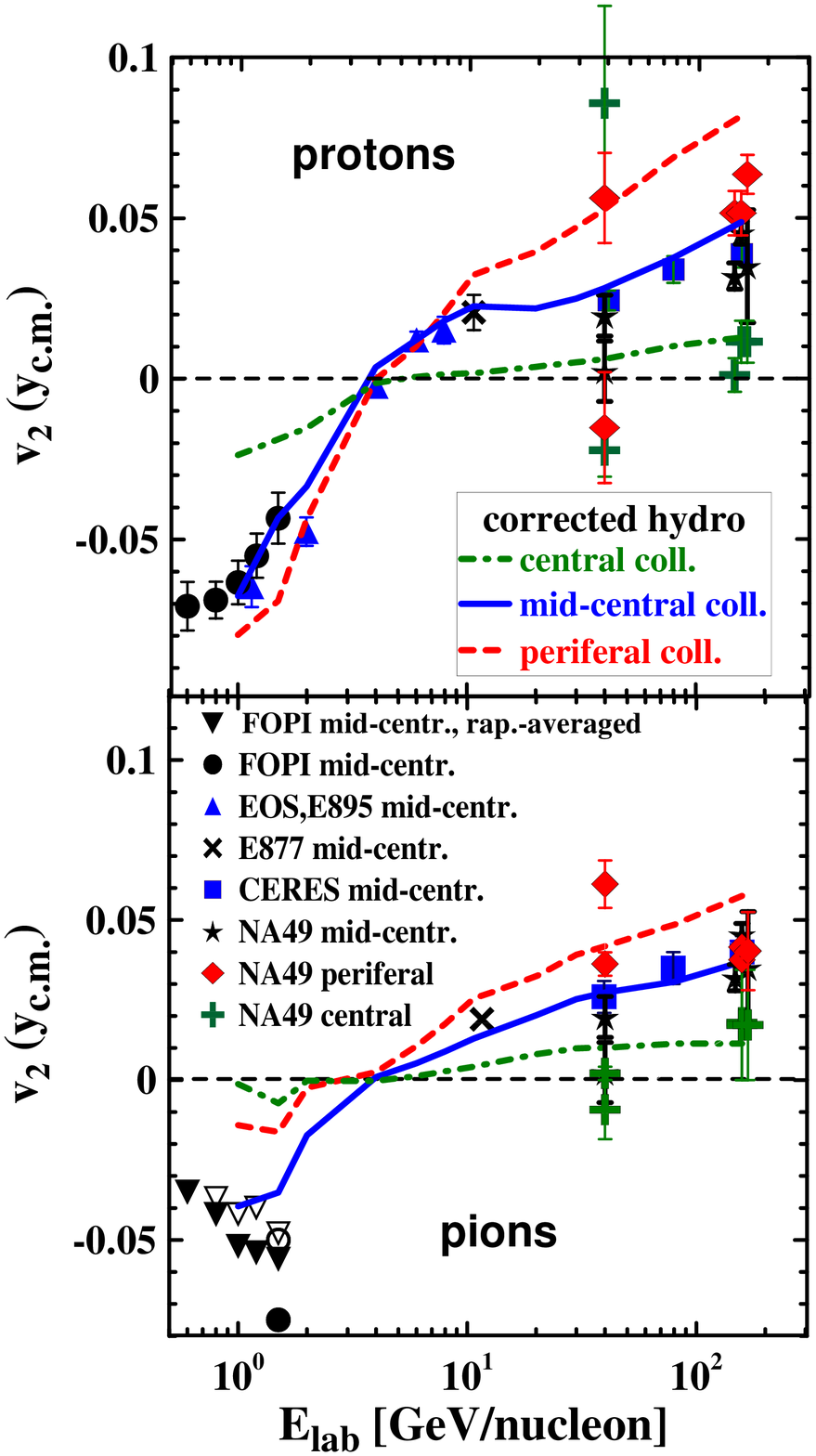}
\caption{(Color online)
Same as Fig. \ref{fig2} but for 3FD results reduced in accordance with
Eqs. (\ref{v2-red}), (\ref{Kn1}) with $c_s\hsp\sigma_{\rm tr}=2.3$~mb. The
results of 3FD calculations
for the intermediate ($K=190$ MeV) EoS and data  are shown for central,
midcentral and peripheral collisions. Different experimental
points (NA49) for the same incident energy and centrality correspond to different
experimental methods of $v_2$ determination, see Ref.~\cite{NA49-03-v1}.}
\label{fig3}
\end{figure}

\section{Elliptic Flow}
\label{Elliptic Flow}

\begin{figure*}[htb]
$\;$\vspace*{-29mm}

\includegraphics[width=15cm]{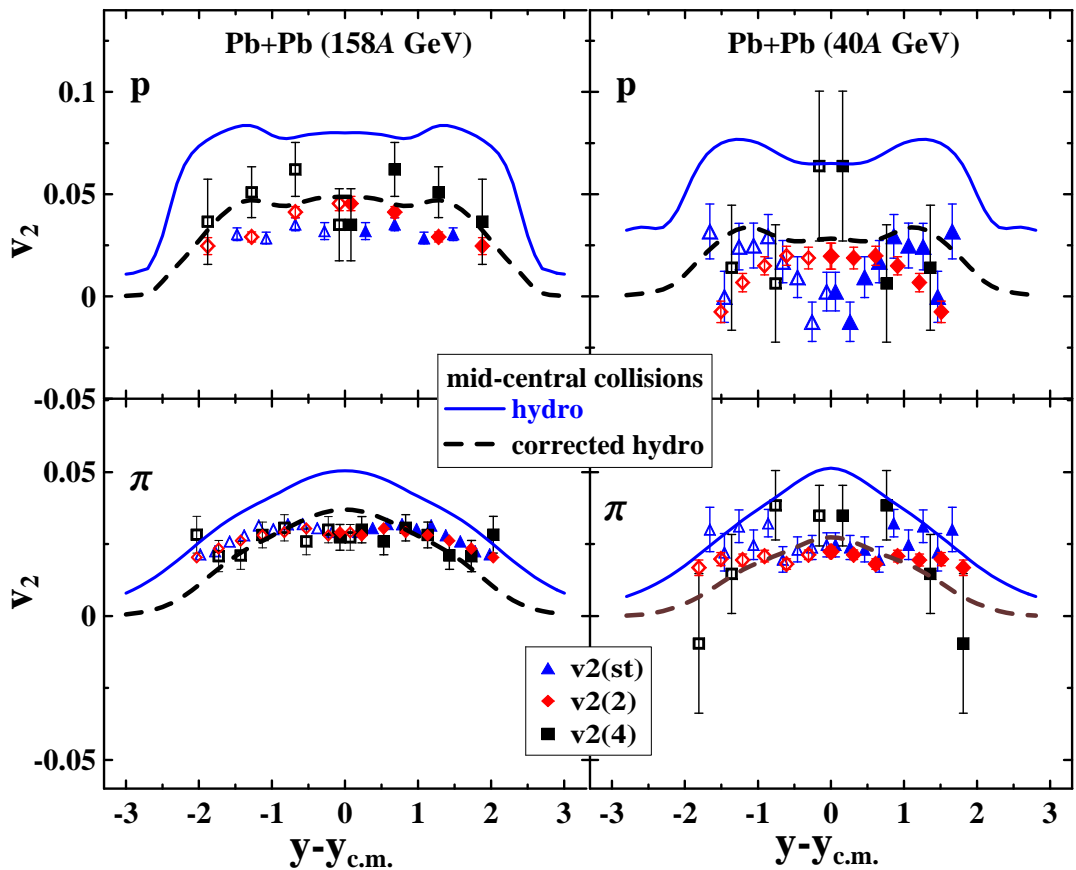}
$\;$\vspace*{-49mm}
\caption{(Color online) Elliptic
flow of protons (upper panels) and charged pions (lower panels) in
mid-central Pb+Pb collisions at $E_{\scr{lab}}=$ 158$A$  GeV (left panels) and
40$A$ GeV (right panels)  as a function of rapidity. The 3FD calculations
are performed at $b=$ 5.6 fm with the intermediate EoS.
Experimental data \cite{NA49-03-v1} obtained by different
methods are displayed: by the standard method ($v_2(\rm st)$)
and by the method of $n$-particle correlations ($v_2(n)$).
Full symbols correspond to
measured data, while open symbols are those reflected
with respect to the midrapidity. Circles in left panels show the updated NA49
data~\cite{NA49-98-v1} with the acceptance 0.05 $<p_T<$ 0.35 GeV/c for pions  and
0.6 $<p_T<$ 2.0  Gev/c for protons.}
\label{fig4}
\end{figure*}

The elliptic flow, defined as $v_2=\langle \cos 2 \phi\rangle$ \cite{Voloshin96},
is the second coefficient in Fourier expansion of the
azimuthal-angle\hsp\footnote{
$\phi$ is the angle with respect to the reaction plane.
}
 dependence of the single-particle distribution function of a hadronic species~$a$,
 $(E\hsp d{^{\hsp 3}}\hspace*{-1pt}N_a/d^{\,3}p)$,
\begin{eqnarray}
 \label{eq-v2}
\hspace*{-4mm}
v_2^{(a)} (y)=
\frac{\displaystyle \int d^{\hsp 2} p_{\,T}  \left[(p^{\hsp 2}_x- p^{\,2}_y)/p^{\hsp 2}_{\,T}\right]
E\hsp d^{\hsp 3}\hspace*{-1pt}N_a/d^{\hsp 3}p }%
{\displaystyle \int d^{\hsp 2} p_{\,T}
E\hsp d^{\hsp 3}\hspace*{-1pt}N_a/d^{\hsp 3}p }\,,
\end{eqnarray}
where $\bm{p}_{\,T}$ is the transverse momentum of the particle, $p_x$ and
$p_y$ are its $x$ and $y$ components. In calculating $v_2$ for pions and
protons we take into account contributions of resonance decays.

Figure \ref{fig2} summarizes the 3FD results
\cite{3FDflow,3FD-GSI07,3FD} for the elliptic flow.
The calculations have been done for Au+Au collisions at $b=$ 6 fm
(SIS and AGS energies)
and Pb+Pb collisions at \mbox{$b=$ 5.6 fm} (SPS energies).
It is seen that
the proton data
are well reproduced at low bombarding
energies,
where the squeeze-out effect dominates. The latter is  caused by
shadowing of expanding participant matter by spectator parts of the
colliding nuclei.
Note that late freeze-out turns out to be preferable for pion data at low energies,
while the reproduction of these data is still
far from being perfect.
At higher energies, when the standard collective mechanism of the
elliptic flow formation starts to work, the 3FD model noticeably
overestimates
both the proton and pion data.
The calculated results only weakly depend on
the stiffness of the EoS.
Tuning the freeze-out condition does not help very much to reduce the disagreement.
The proton elliptic flow turns out to be quite insensitive to this condition.
%
The sensitivity of the pion elliptic flow is higher. However, even very early
freeze-out, which looks preferable for pions, does not allow to fit the data.
The different sensitivity of the proton and pion $v_2$ to the freeze-out condition
is a consequence of three-fluid nature of our model. Saturation of $v_2$ in the
fireball fluid occurs later than that in the baryon-rich fluids. Since contribution
of the fireball fluid into the pion yield is larger than for protons,
the saturation of the total pion $v_2$ happens also later.
%
%
Variation of the inter-fluid friction does not improve agreement with
the data either, while destroying the description of other observables.

The calculations show that for
$\varepsilon_{\rm frz}\lesssim 0.2$ GeV/fm$^3$,  both the proton and pion elliptic
flows stay practically unchanged. Therefore, the results obtained for
$\varepsilon_{\rm frz}=0.2$ GeV/fm$^3$ can be naturally associated with hydrodynamic
limit\hsp\footnote{Note that these results correspond to the  3FD initial conditions.}
of the elliptic flow.

It is natural to associate the overestimation of experimental $v_2$
values with dissipative effects during the
expansion and freeze-out of the participant matter. To take these effects into
account we
use an empirical formula suggested in~\cite{Ollitrault05}:
\begin{eqnarray}
 \label{v2-red}
v_2 = \frac{v_2^{\rm hydro}}{1+{\rm Kn}/{\rm Kn}_0}\,.
\end{eqnarray}
Here $v_2$ is the observed value of the elliptic flow, $v_2^{\rm
  hydro}$ is its hydrodynamic limit, ${\rm Kn}$ is an effective
Knudsen number defined as
\begin{eqnarray}
 \label{Kn}
{\rm Kn} =  \frac{\lambda}{R}\,,
\end{eqnarray}
where $\lambda$ is a mean-free path of a particle and
$R$ is a characteristic size of the system (e.g., the radius of the nucleus),
${\rm Kn}_0\sim 1$ is a constant. The recent transport calculation in two spatial
dimensions~\cite{Ollitrault08} resulted in ${\rm Kn}_0\simeq 0.7$.
We use this value of ${\rm Kn}_0$ for our estimates below.

As argued in Ref. \cite{Ollitrault05}, the Knudsen number can be
represented in the form
\begin{eqnarray}
 \label{Kn1}
\frac{1}{{\rm Kn}} \simeq  \frac{c_s\hsp\sigma_{\rm tr}}{4S} \frac{dN_{\rm tot}}{dy}\,.
\end{eqnarray}
Here $dN_{\rm tot}/dy$ is the total (charged plus neutral) hadron
multiplicity per unit rapidity, which equals approximately 3/2 of the charged-hadron
rapidity density $dN_{\rm ch}/dy$, $\sigma_{\rm tr}$ is the transport cross section,
$c_s$ is the sound velocity in the medium, and $S$ is the
transverse overlap area between two colliding nuclei. The latter is defined as
$S=\pi \sqrt{\langle x^2\rangle  \langle y^2\rangle}$
\cite{Jacobs00}, where $\langle x^2\rangle$ and $\langle y^2\rangle$
are mean values of $x^2$ and $y^2$ over the overlap zone\footnote{We follow
the conventional definition of
$S$  \cite{Jacobs00} which differs from
that in Refs. \cite{Ollitrault05,Ollitrault08} by the factor of 4.}
defined by the collision geometry.

Similarly to Ref. \cite{Dumitru07}, we fit the data displayed in
Fig.~\ref{fig2} by using \mbox{Eq.~(\ref{v2-red})} with ${\rm Kn}$ defined by~(\ref{Kn1})
and $v_2^{\rm hydro}$ taken from the 3FD calculation with the late
freeze-out, i.e. at $\varepsilon_{\rm frz}=0.2$ GeV/fm$^3$. The rapidity density
$dN_{\rm ch}/dy$ is also calculated within the 3FD model.
Note that the charged-hadron rapidity distributions
are well reproduced by the 3FD model (see Fig.~\ref{fig1}).
The reduction factor $(1+{\rm Kn}/{\rm Kn}_0)^{-1}$ 
is applied only at
$E_{\rm lab} > 4$A GeV, since it is not justified
at lower energies due to 
importance of the squeeze-out effects.
Indeed, Eq. (\ref{v2-red}) was deduced from simulations of unbiased expansion of a system
into the transverse directions \cite{Ollitrault05}. In this case the elliptic flow can
be only positive. The
squeeze-out means that the transverse expansion is screened by spectators. The latter results
in a suppressed and even negative elliptic flow.
Therefore, at lower incident energies purely hydrodynamical results for the late freeze-out
($\varepsilon_{\rm frz}=0.2$ GeV/fm$^3$) are presented.
Probably, the elliptic flow in the
squeeze-out region should be corrected for the dissipative effects but certainly not by
means of Eq. (\ref{v2-red}).

The results of our fit are presented in Fig. \ref{fig3}. One can see
that a good reproduction of the AGS and SPS data is achieved with the help of
a single fitting parameter $c_s\hsp\sigma_{\rm tr}\simeq 2.3$ mb. For the
midcentral collisions this leads to the estimate ${\rm Kn}\sim 0.7$ at the midrapidity.
Introduction of the dissipative correction also substantially improves the
rapidity dependence of the elliptic flow. This is indeed seen from
Fig. \ref{fig4}. Taking $c_s^{\hsp 2}\sim 0.15$~\cite{Ch05}, we arrive at the estimate
$\sigma_{\rm tr} \sim 6$ mb for top AGS and SPS energies.
This is a surprisingly low value of the cross section. However, taking
into account that $\sigma_{\rm tr}$ is the transport cross section (which is, in
general, lower than the total one) and having in mind uncertainties in the Knudsen number
definition (\ref{Kn1}), this value seems to be still acceptable.

\begin{figure}[thb]
\hspace*{-7mm}
\includegraphics[width=9cm]{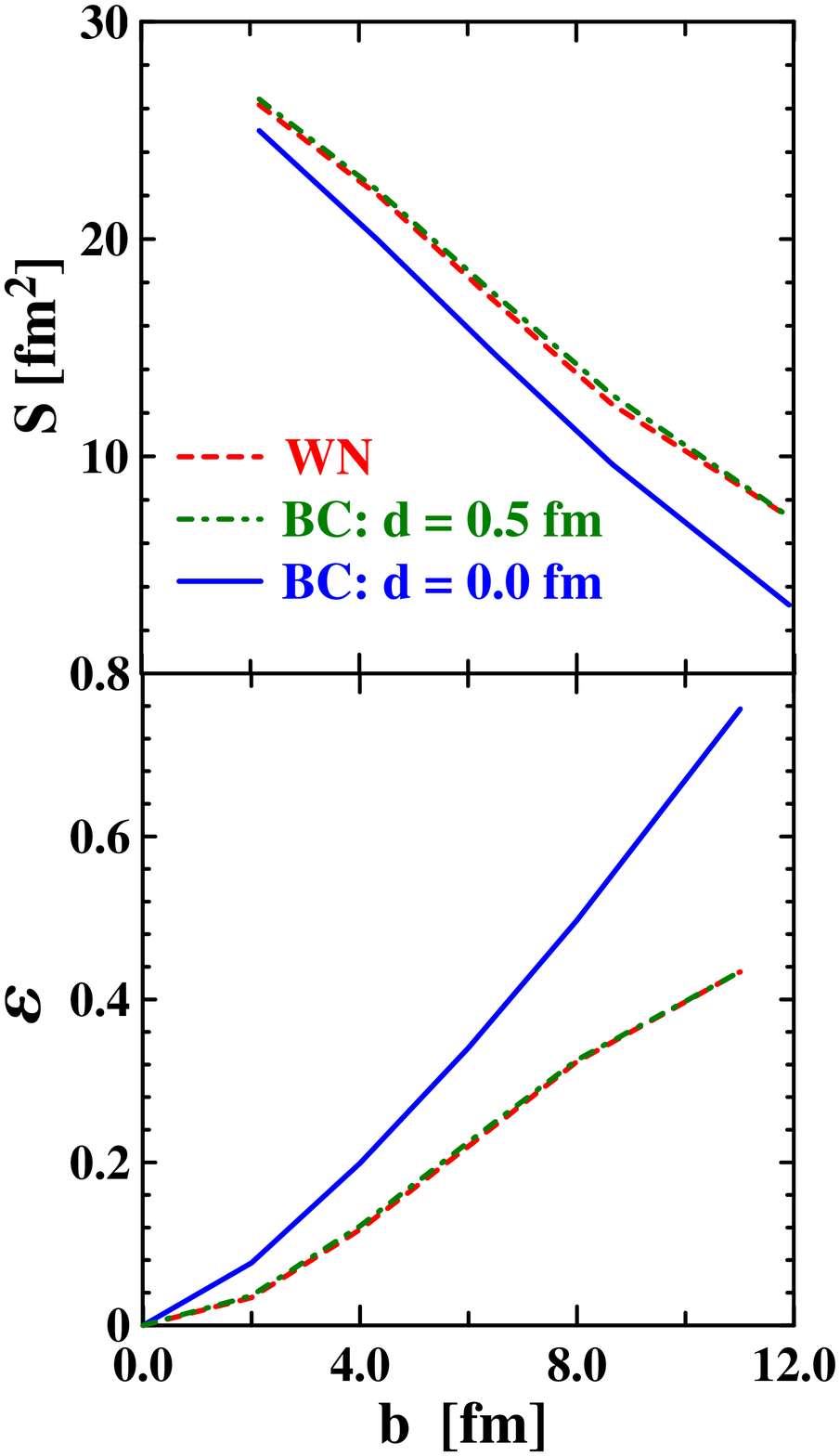}
$\;$\vspace*{-9mm}
\caption{(Color online)
The transverse overlap area $S$ and spatial eccentricity
$\varepsilon$ as functions of impact parameter in Au+Au collisions
for different surface diffusenesses of the Au nucleus ($d$) and
different weights of averaging: the wounded-nucleon (WN) and
the binary-collision (BC) weights~\cite{Jacobs00}.
}
\label{fig5}
\end{figure}
\begin{figure}[thb]
\hspace*{-13mm}
\includegraphics[width=10.5cm]{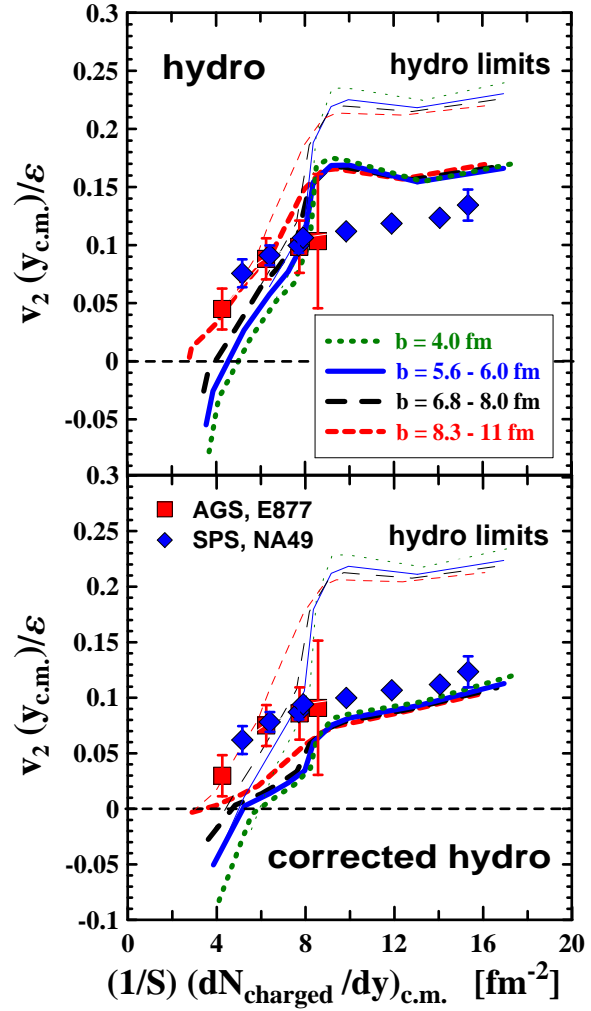}
$\;$\vspace*{-9mm}
\caption{(Color online)
Pion elliptic flow at midrapidity divided by the eccentricity as a
function of the charged-hadron rapidity density in
Au+Au (AGS energies) and Pb+Pb (SPS energies) collisions at
different centralities.
Bold lines in upper and bottom panels display, respectively, the
hydrodynamic $v_2$ calculated with the standard
freeze-out energy density  ($\varepsilon_{\rm frz}=0.4$ GeV/fm$^3$)
and the hydrodynamic $v_2$ corrected accordingly to
Eq. (\ref{v2-red}). Thin lines represent the uncorrected hydrodynamic
calculation with the late freeze-out ($\varepsilon_{\rm frz}=0.2$ GeV/fm$^3$).
The 3FD calculations have been done with the intermediate EoS
($K=190$ MeV). Compilation of experimental data is from
Ref. \cite{Voloshin08}.
}
\label{fig6}
\end{figure}

\begin{figure}[thb]
$\;$\vspace*{-19mm}

\includegraphics[width=8.1cm]{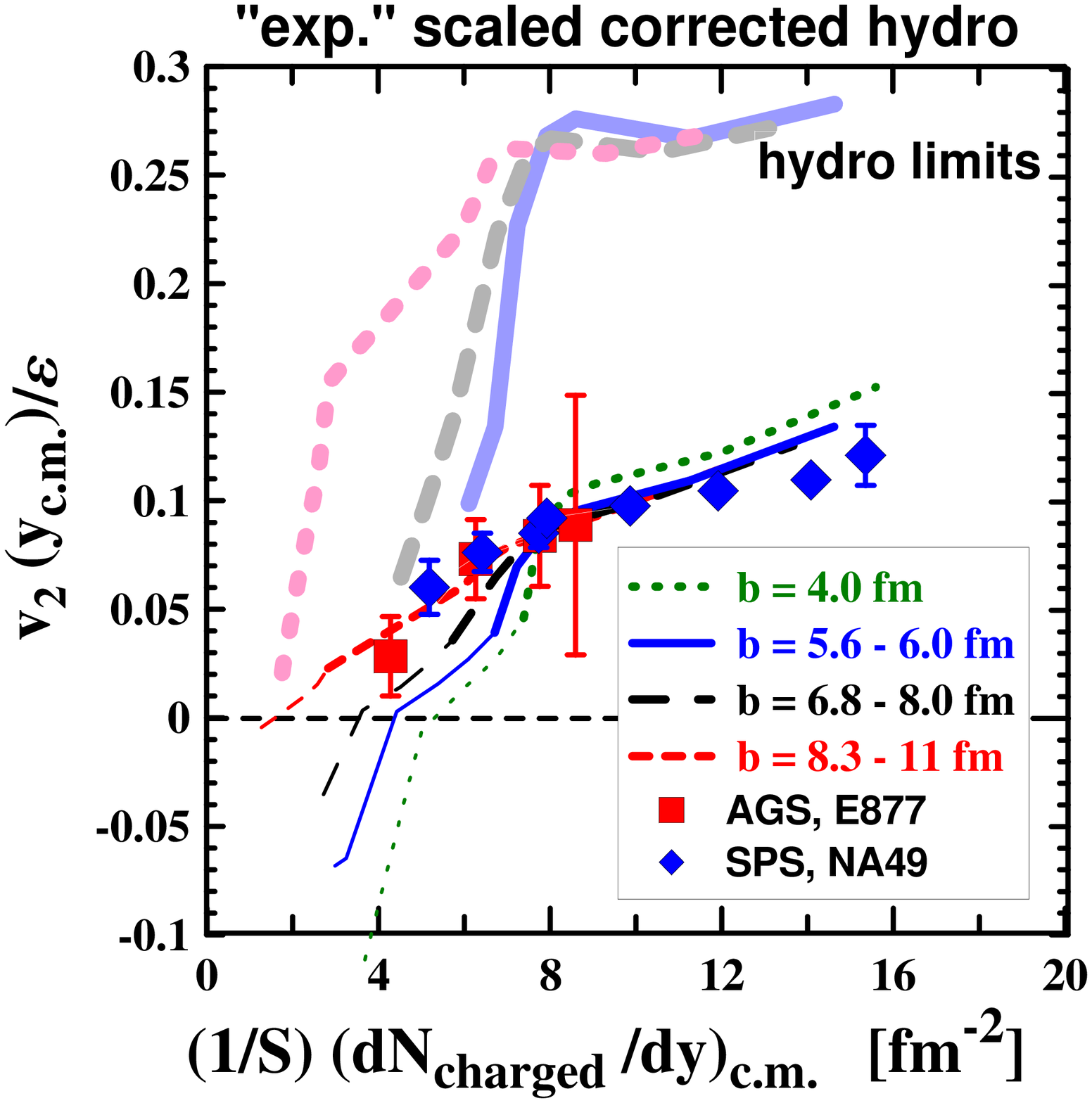}
$\;$\vspace*{-19mm}
\caption{(Color online)
The same as in bottom panel of Fig. \ref{fig5} but scaled with
``experimental'' scaling quantities $S$ and $\varepsilon$.
Elliptic flow at incident energies $E_{\rm lab} \geq 10A$ GeV is
displayed by bold lines, while that at lower energies
$E_{\rm lab} \leq 10A$ GeV -- by thin lines.
}
\label{fig7}
\end{figure}

Let us discuss now an approximate scaling behavior of
$v_2$ proposed in Refs.~\cite{Voloshin08,Voloshin00}.
This behavior is observed when $v_2$ scaled with the initial
eccentricity $\varepsilon$ is plotted as a function of
$dN_{\rm ch}/dy$ scaled with the cross section of the nuclear overlap $S$.
Both $\varepsilon$ and~$S$ are determined by the collision geometry.
In particular, $\varepsilon$ is defined as
\begin{eqnarray}
 \label{eps}
\varepsilon = \frac{\langle y^2\rangle - \langle x^2\rangle}%
{\langle y^2\rangle + \langle x^2\rangle}\,\,.
\end{eqnarray}
Below we calculate $\langle x^2\rangle$ and $\langle y^2\rangle$
with either the wounded-nucleon~(WN) or the binary-collision~(BC) weights,
for details see Ref. \cite{Jacobs00}. These calculations are
based on the usual Woods--Saxon profile of nuclear density.
Within the 3FD model the initial nuclei are represented by sharp-edged
spheres. To be consistent with the model, we first
calculated $\varepsilon$ and $S$ using the Woods--Saxon parametrization
with zero diffuseness ($d=0$) and the
BC weight.  The obtained results are shown in Fig.~\ref{fig5} by the solid lines.

Our scaling analysis is summarized in Figs.~\ref{fig6} and \ref{fig7}.
First of all, we have found that there is no scaling for proton
elliptic flow: points corresponding to different energies and impact
parameters populate a relatively broad band rather than a universal line.
This is not surprising, since scaling either with the BC or WN weights is
inappropriate for nucleons. Indeed, their rapidity spectra are strongly constrained
by the conservation of the baryon number. As a result, contrary to pions,
the number of participating nucleons is determined mainly by the initial
geometry and it is not proportional to the number of binary collisions.

At the same time the pionic $v_2$ exhibits an approximate scaling
behavior already for pure hydrodynamic calculation, see the top panel
of Fig. \ref{fig6}.
This scaling is more pronounced at higher densities of charged
particles, i.e. at higher incident energies. However, this scaling
still substantially differs from the observed one,
cf. experimental points in Fig. \ref{fig6}. When we correct the
hydrodynamic results according to Eq. (\ref{v2-red}) (see the bottom
panel of Fig. \ref{fig6}), the resulting~$v_2$ reveals a scaling
behavior which is closer to the experimental results.
By thin lines in Fig.~\ref{fig6} we also show the hydrodynamic limits
of the elliptic flow. The latter are obtained from the 3FD calculation
with the late freeze-out ($\varepsilon_{\rm frz}=0.2$~GeV/fm$^3$).

Although the scaling factors  $\varepsilon$ and $S$, used in
Fig. \ref{fig6} (see the ``BC: d = 0.0'' curves in Fig.~\ref{fig5}), are
consistent with assumptions of the 3FD model, they
differ from those used for scaling of the experimental data
\cite{Voloshin08}. The experimental points in Figs.~\ref{fig6} and \ref{fig7}
are scaled with $\varepsilon$ and $S$ calculated with the WN weight and
$d=0.535$ fm \cite{Jacobs00} (see the ``WN'' curves in Fig.~\ref{fig5}).
It turns out that $\varepsilon$ and $S$
calculated with the BC weight and
$d=0.5$ fm agree fairly well with the ``experimental'' scaling factors.

Upon applying the ``experimental'' scaling factors to the corrected
elliptic flow, we obtain the scaling-like behavior displayed in
Fig. \ref{fig7}. Here the experimental scaling turns out to be well
reproduced in its high-density part, while the calculated
elliptic flow still reveals no scaling at low charge densities.
Note, however, that the lowest incident energy, at which the pion
elliptic flow
have been measured, is  $E_{\rm lab} = 11.5A$ GeV (the E877 data
\cite{E877-v2} in \mbox{Fig.~\ref{fig7})}.
If we consider $v_2$ only at high energies  $E_{\rm lab} \gtrsim 10A$ GeV
(lower bold lines in \mbox{Fig.~\ref{fig7})}, the reproduction of the scaling
behavior becomes much better.
In principle, if we let the fitting parameter $c_s\hsp\sigma_{\rm tr}$ vary
with the incident energy (i.e. smoothly join the corrected $v_2$ at high
energies with those kept unchanged at low energies), then the scaling at
$E_{\rm lab} \gtrsim 10A$ GeV can be
better reproduced. However, we deliberately avoid such a
multi-parameter fit in order to keep the underlying physics more
transparent.

At lower AGS energies the 3FD model certainly predicts no
scaling. Apparently, this is a consequence of the partial shadowing of
the transverse expansion by spectators. At lowest AGS energies this
shadowing becomes dominant and leads to the negative elliptic
flow.

It should be mentioned that scaling factors taking into account
fluctuations of the initial eccentricity lead to a better quality of the experimental
scaling \cite{Voloshin07} (at least at RHIC energies).
Since the AGS-SPS data analyzed here were obtained
without applying the fluctuation corrections, we also do not use them in
the scaling factors. Another aspect is influence of initial-state
fluctuations on the hydrodynamic results. In our calculation we do not introduce
such fluctuations. In Ref. \cite{Hama09} it was shown
that the initial-state fluctuations can noticeably affect the elliptic flow
in semicentral collisions,
especially at high transverse momenta and marginal rapidities.
However, for the midrapidity region in midcentral collisions, considered here,
the effect of fluctuations is quite moderate.

\section{Dissipation }

Having determined the effective Knudsen number, we can now estimate the
role of dissipative effects during the expansion stage of a nuclear collision.
Let us first attribute all this dissipation to the fluid viscosity.
To estimate the latter,
we use nonrelativistic formulas having in mind that the order of magnitude of
hadronic masses is $m \sim 1$ GeV while the freeze-out temperature $T$ is of the
order of 100~MeV. Note that even at the top SPS energy, where the pion yield
almost twice exceeds the yield of all other particles, the system at
the freeze-out stage consists mainly of heavy baryon and meson resonances
which only later decay into pions. At this stage, thermal pions comprise
less than 50\% of all observed pions.

From the kinetic theory of simple gases
one can estimate the shear viscosity coefficient as~\cite{Dan85}
\begin{eqnarray}
 \label{kin-visc}
\eta \simeq\frac{1}{3}\hsp\langle p \rangle\hsp n\hsp\lambda\, ,
\end{eqnarray}
where $\langle p \rangle\simeq\sqrt{8mT/\pi}$ is an average thermal momentum of a
typical hadron, $\lambda=(n\sigma_{\rm tr})^{-1}$ is its transport mean free path and
$n$~is the total particle density of the medium\hsp\footnote{The calculation of Ref.
\cite{Lif81} for the hard-sphere gas gives a
similar expression for $\eta$ with $\sigma_{\rm tr}=4\pi r_0^2$, where $r_0$ is the
radius of the sphere.}.
For the rough estimate we neglect the angular anisotropy of
hadron-hadron cross sections.
Taking $\sigma_{\rm tr}=6$\,mb, obtained from the fit of $v_2$ data in Sect.~III,
we get the estimate
\bel{etest}
\eta\simeq (1.4- 1.7)~{\rm fm}^{-3}\,,
\ee
for $T=(0.1- 0.15)$\,GeV. One can estimate the role of thermal pions
(not hidden in resonances) by using the
relativistic generalization of the above expressions suggested in Ref.~\cite{Gor08}.
The direct calculation shows that pions contribute to $\eta$ not
more than 30\% in the same domain of $T$.

The relative strength of dissipative effects in fluid dynamics can be characterized by the
ratio of $\eta/s$ where $s$ is the entropy density. The latter quantity can be estimated
by using the Sackur--Tetrode formula~\cite{Land-Lif}
\bel{s}
s = n \ln \left[\left(\frac{m\hsp T}{2\hsp\pi}\right)^{3/2}\frac{1}{n} \right]
+ \frac{5}{2}n\,.
\ee
For $T=(0.1- 0.15)$~GeV and $n=0.2$ fm$^{-3}$ this formula leads to the estimate
$s\simeq (0.6- 0.7)$~fm$^{-3}$.
The same calculation at $n=0.4$ fm$^{-3}$ gives higher values $s\simeq (0.8- 1.1)$~fm$^{-3}$.
However, Eq.~(\ref{s}) corresponds to the gas of identical particles and, therefore,
underestimates the entropy. The calculations for the gas of hadronic resonances
with excluded volume corrections~\cite{Sat08} gives entropy densities
which are by about a factor of two higher. Finally, in the considered domain of $T$ and $n$
we get the estimate $\eta/s\sim 1 - 2$\,.
At lower incident energies, noncentral rapidities and/or for more peripheral
collisions the $\eta/s$ ratio may be even higher. This is expected, because the ideal hydrodynamics
becomes less justified in these regions.

The obtained value of $\eta/s$ is of the order of minimal values
observed in water and liquid nitrogen~\cite{Kapusta06}. Therefore, the
strongly-interacting matter in the considered energy range indeed behaves like a
viscous liquid rather than a gas.
However, the obtained large values of the Knudsen number
${\rm Kn} \sim 0.7$ may be interpreted as if the fluid dynamics is not
applicable at all.
This would be indeed so, if the estimated viscosity corresponded to
entire stage of the hydrodynamic expansion. In fact, the estimated
Knudsen number and $\eta/s$ ratio should be also attributed to late stages of
the expansion after the freeze-out, i.e. to the afterburner.
During this late stage the hydrodynamics becomes
inapplicable and hence it is not surprising that the resulting effective
Knudsen number turns out to be so large. This implies that Knudsen
numbers corresponding to the hydrodynamic expansion
may be still small, making hydrodynamics applicable.
The argument in favor of such an interpretation is that the
3FD model is able to reproduce a large body of experimental
data in the AGS-SPS energy range \cite{3FD}.
Another evidence in favor of this interpretation follows
from the study~\cite{Hirano07,Bleicher09} of a post-freeze-out evolution
which has shown that the afterburner is responsible for
the major part of the discrepancy between the ideal hydro
predictions and experimental data for $v_2$.
 Therefore, the above estimated values of $\eta/s$
should be considered as an upper
limit for of the hydrodynamic expansion.

The above arguments are not only relevant to the AGS-SPS energy
domain. A similar analysis~\cite{Dumitru07} of the PHOBOS data
at RHIC energies also gives rather large values of the Knudsen number.

%
%

\section{Discussion and Conclusions}

In this paper we have analyzed the elliptic
flow in the SIS-AGS-SPS energy range within the 3FD model.
Direct hydrodynamical calculations of the elliptic flow result in a
good reproduction of experimental data at
SIS and lower AGS energies
(description of the FOPI pion data is still far from being perfect)
but
considerably overestimate the data at top AGS and SPS energies.
The latter problem cannot be cured by neither variation of the EoS
stiffness nor the freeze-out criterion.
Changing the inter-fluid friction forces does not solve this problem either.
In this paper we
attribute this problem to dissipative effects during the expansion and
freeze-out stages. In order to estimate the role of dissipation
from the difference between the 3FD results and the observed data,
we apply a simple formula, proposed in Ref. \cite{Ollitrault05}, where
this difference is expressed in terms of the Knudsen number.
It is shown that the interpretation of the disagreement between the 3FD
model and experimental $v_2$ data due to dissipation turns out to be
fruitful. With the help of a singe parameter we are able to fit
the calculated $v_2$ values to the observed data in a broad range of
incident energies, centralities and rapidities.
Moreover, the experimental scaling behavior of pion elliptic
flow scaled with the initial eccentricity versus  charged-hadron-multiplicity
density per unit transverse area turns out to be reasonably reproduced.

Proceeding from values of the Knudsen number, deduced from the above fit,
we estimated the viscosity-to-entropy ratio as $\eta/s \sim 1-2$.
These values of $\eta/s$ are of the order of minimal values
observed in water and liquid nitrogen \cite{Kapusta06}. Therefore, the
nuclear matter in the considered energy range indeed behaves like a
liquid, however, not so perfect as at RHIC energies.

The estimated $\eta/s$ ratio accumulates the effects of dissipation
at the hydrodynamic expansion stage
and the afterburner stage after the hydrodynamic freeze-out.
Having in mind that the afterburner may give an important contribution
to dissipation we conclude that the estimated $\eta/s$ values represent only an upper
limit for this quantity when applied to the hydrodynamic expansion stage
until the kinetic freeze-out.
On the other hand, the authors of Ref.~\cite{Gor08} have estimated the $\eta/s$
ratios on the chemical freeze-out line observed for central collisions of heavy nuclei
at beam energies from SIS to RHIC. Within the excluded-volume
hadron-resonance-gas model, they found that $\eta/s$ are larger than 0.3-0.5
depending on the hard core radius. We would like to stress that these values should
be lower than ours because they correspond to the earlier chemical freeze-out stage.

Recently the CERES collaboration reported that $\eta/s = 0.021 \pm 0.068$  at the top
SPS energy \cite{CERES09}. This result is certainly in contradiction to our estimate of
this ratio. In Ref. \cite{CERES09} the $\eta/s$ ratio was deduced from analysis of
two-pion correlation measurements, more precisely, of the transverse-momentum
dependence of the longitudinal pion source radius. The analysis was performed
on the basis of boost-invariant (Bjorken) solution to the relativistic viscous
hydrodynamics. Concerning the latter we would like to mention
that it is not well justified experimentally since the pion rapidity spectra
are far from being flat at SPS energies. As it is illustrated in Ref. \cite{3FDfrz},
the 3-fluid dynamics of nuclear system at 158$A$ GeV is quite different from the
Bjorken picture. This could be a reason of disagreement between
our results and the results of Ref.~\cite{CERES09}.

\vspace*{5mm} {\bf Acknowledgements} \vspace*{5mm}

We are grateful to S. Voloshin for clarifying the meaning of the
elliptic-flow scaling.
This work was supported in part by
the Deutsche Bundesministerium f\"ur Bildung und Forschung (BMBF
project RUS 08/038),  the  Deutsche
Forschungsgemeinschaft (DFG project 436 RUS \mbox{113/957/0-1}), the
Russian Foundation for Basic Research (RFBR grant 09-02-91331) and the
Russian Ministry of Science and Education (grant NS-3004.2008.2).
This work was also partially supported by the Helmholtz International
Center for FAIR within the framework of the LOEWE program
(Landesoffensive zur Entwicklung Wissenschaftlich-Okonomischer
Exzellenz) launched by the State of Hesse.

\end{document}